\documentclass[aps,superscriptaddress,altaffilletter,lengthcheck,tightenlines,showpacs,showkeys]{article}
\usepackage{graphicx}
\usepackage{subcaption}
\pdfoutput=1
\newcommand{\ben}{\begin{eqnarray}}
\newcommand{\een}{\end{eqnarray}}
\newcommand{\be}{\begin{equation}}
\newcommand{\ee}{\end{equation}}
\newcommand{\ba}{\begin{eqnarray}}
\newcommand{\ea}{\end{eqnarray}}
\newcommand{\n}{\label}

\newcommand{\ga}{\gamma}
\newcommand{\ro}{\rho}

\newcommand{\dark}{\rm{Dark}}
\usepackage{amsmath,amssymb,amsthm,amscd}
\usepackage[dvipdf]{epsfig}
\usepackage{color}
\usepackage[colorlinks,citecolor=blue]{hyperref}

\begin{document}
\title{Non-linear coupling in the dark sector as a running vacuum model\\}

\author{
Josue De-Santiago\thanks{Instituto de Ciencias F\'isicas, Universidad Nacional Aut\'onoma de M\'exico. josue@fis.unam.mx} \,, 
Iv\'an E. S\'anchez G.\thanks{Depto. de F\'isica, FCEyN and IFIBA, Universidad de Buenos Aires, Buenos Aires, Argentina isg.cos@gmail.com and isg@df.uba.ar.} \, 
and David Tamayo\thanks{Departamento de F\'{i}sica, Centro de Investigaci\'{o}n y de Estudios Avanzados del IPN, AP 14-740, Ciudad de M\'{e}xico 07000, M\'{e}xico. dtamayo@fis.cinvestav.mx}}
\date{}
\maketitle

\begin{abstract}
In this work we study a phenomenological non-gravitational interaction between dark matter and dark energy. The scenario studied in this work extends the usual interaction model proportional to the derivative of the dark component density adding to the coupling a non-linear term of the form $Q = \rho'/3(\alpha + \beta \rho)$. This dark sector interaction model could be interpreted as a particular case of a running vacuum model of the type $\Lambda(H) = n_0 + n_1 H^2 + n_2 H^4$ in which the vacuum decays into dark matter. For a flat FRW Universe filled with dark energy, dark matter and decoupled baryonic matter and radiation we calculate the energy density evolution equations of the dark sector and solve them. The different sign combinations of the two parameters of the model show clear qualitative different cosmological scenarios, from basic cosmological insights we discard some of them. The linear scalar perturbation equations of the dark matter were calculated. Using the CAMB code we calculate the CMB and matter power spectra for some values of the parameters $\alpha$ and $\beta$ and compare it with $\Lambda$CDM. The model modify mainly the lower multipoles of the CMB power spectrum remaining almost the same the high ones. The matter power spectrum for low wave numbers is not modified by the interaction but after the maximum it is clearly different. Using observational data  from Planck, and various galaxy surveys we obtain the constraints of the parameters, the best fit values obtained are the combinations $\alpha = (3.7 \pm 7 )\times 10^{-4} $, $-(1.5\times10^{-5} {\rm eV}^{-1})^{4} \ll \beta < (0.07 {\rm eV}^{-1})^4$.
\end{abstract}
\vskip 1cm


\section{Introduction}
The Supernovae Type Ia (SNIa) was the first observational evidence that the Universe is in a period of accelerated expansion \cite{Riess1} \cite{Perlm}. This success has been confirmed by other observations such as the Cosmic Microwave Background radiation (CMB) anisotropy spectrum \cite{Adam:2015rua}, the Baryon Acoustic Oscillations (BAO) \cite{BAO}, among others. To explain the accelerated expansion of the Universe within the framework of General Relativity it has been postulated the existence of a unknown component with negative pressure called dark energy. The simplest dark energy model is to assume that it is the (positive) cosmological constant $\Lambda$ of the Einstein equations. It is the dominant component of the Universe energy density but its particle physics origin is an open problem.

Similarly, observations like galactic rotation curves, gravitational lensing, galaxy clusters, large-scale cosmological structure \cite{Drees} \cite{Garrett}, and CMB \cite{WMAP} \cite{Planck1} \cite{Planck2} indicate the presence of another mysterious Universe component called dark matter which main features are to be gravitationally attractive and do not absorb or emit electromagnetic radiation. Despite dark matter is also an important component contributing with around the $25\%$ of the energy density of the Universe and there is an ample evidence of its gravitational effects, its  nature from particle physics remains unclear.

Summarizing, data show that the Universe is filled with two non-baryonic components (dark energy and matter) which represent the $\sim 95\%$ of the total energy density. The cosmological model considering dark matter as a non-relativistic fluid, the so-called Cold Dark Matter (CDM), and the $\Lambda$ as the dark energy is commonly knows as $\Lambda$CDM model. This model is widely successful, predicting the observable Universe and supported by observations of different type. However, the $\Lambda$CDM model has important conundrums like the cosmological constant problem \cite{Weinberg1989}. Basically when is considered the cosmological constant as dark energy, it imply that the observed vacuum energy density in cosmology is around 100 orders of magnitude lower then the calculations done by quantum field theory. Also, if we interpret dark energy (cosmological constant) as a material fluid as the other known components (photons, baryons, neutrinos etc.) it has a constant energy density ($\rho_{\text{DE}} = \Lambda/8\pi G$) without dynamics.

To tackle these problems (or part of them) has been proposed many extensions, and alternatives to the $\Lambda$CDM model like MOND, brane cosmology, SUGRA cosmology, $f(R)$ gravity among others. One of these alternatives are decaying (or running) vacuum models which expect that $\Lambda$ (the vacuum) should not be strictly constant but it is a smooth functions of some dynamical cosmological quantity such like the Hubble function $H(t)$ \cite{Grande}, \cite{Alcaniz}. The main idea in these class of models is to assume that the vacuum is a dynamical material component that evolves (as the other components) with the time, in order to preserve the condition energy-momentum tensor conservation $\nabla_\nu T^{\mu\nu}=0$, it is imposed an interaction of the vacuum with the other fluids of the Universe i.e. it does not conserve individually. It were first proposed from phenomenological arguments but later was grounded by the Renormalization Group theory \cite{Sola}. Their global cosmological consequences have been studied in \cite{Perico}, their observational viability in \cite{Gomez-Valent}, primordial gravitational waves and running vacuum in \cite{Tamayo} and lineal scalar perturbations in this context \cite{Perico:2016kbu}. Other models that propose the dark energy as a dynamical scalar field, as the quintessence model \cite{Ferreira} \cite{Stein} \cite{Sahni} is some cases can be reduced to decaying vacuum.

Another important alternative to these models is the possibility of interaction  between the  dark  sector (dark energy and dark matter) \cite{Luis}. The existence of an interaction between dark matter and dark energy it is also possible at theorical level when coupled scalar fields are considered \cite{Amendola1} \cite{Amendola2} \cite{Zimdahl} \cite{Koivisto} \cite{Ziaeepour}.
Models with non-gravitational interaction between dark matter and dark energy have called attention in the last decade. The nature of both dark energy and matter are still unknown, and in principle, the additional interaction between them is possible. This connection between components changes the background evolution of the dark sector \cite{Luis7} \cite{Sanchez} \cite{Marachlian} \cite{Pavon} \cite{Vali}, giving rise to a rich cosmological dynamics compared with non interacting models. On the observational side, the cosmological analysis from Planck data showed that the standard $\Lambda$CDM model remains an excellent fit to the CMB data \cite{Planck2}. However the results of the Hubble constant from Planck are in disagreement with the direct measurements of $H_0$ by the Hubble Space Telescope \cite{Riess3}. There is also some tension in the measurements of $\sigma_8$, the amplitude of the linear power spectrum on the scale of 8 Mpc $h^{-1}$ \cite{Planck3}. There are some interaction models that could alleviate this tension presented in the $\Lambda$CDM model, see \cite{Wang}, \cite{Murgia}. That is why these phenomenological models could be a viable option in the description of the Universe, compatible with a large host of cosmological data.

In this article, we are going to explore an alternative interaction in the dark sector, which is a non linear interaction in the derivative of the dark energy density. We will study how this interaction affects the background evolution along with its impact on the perturbation equations. We are going to constrain the parameters of the model and the cosmic set of parameters by using the CMB data and Large Scale Structure (LSS) observations. For this purpose, we are going to modify the COSMOMC package \cite{Lewis:2002ah}. In section 2 we show the main features and equations of the model proposed as well the evolution of the energy density of the dark sector. In section 3 we write the linear scalar perturbation equations and growth factor, considering the dark sector as background. Next in section 4 we make a comparison of the model parameters with observations, and finally in section 5 we discuss our results and conclusions.

\section{Interaction Model}\label{sec:model}

In the Interaction Scenario, a spatially flat ($k=1$ or $\sum \Omega_i=1$) isotropic and homogeneous universe described by Friedmann-Robertson-Walker (FRW) spacetime is considered. The Universe is filled with four components: radiation, baryonic matter, and two fluids that interact in the dark sector. The first and the second one are decoupled components. The evolution of the FRW universe is governed by the Friedmann and conservation equations,
\be
\label{E1a}
 \frac{3H^{2}}{8\pi G}=\ro_T=\rho_{r}+\rho_{b}+\ro_{m}+\ro_{x},
\ee
\be
\label{E1c}
\dot\ro_{r}+3H\ga_{r}\ro_{r}=0, \qquad \dot\ro_{b}+3H\ga_{b}\ro_{b}=0,
\ee
\be
\label{E1b}
\dot{\ro}_{m}+\dot{\ro}_{x}+3H(\ga_m\rho_{m}+\ga_x\rho_{x})=0,
\ee
where $H=\dot a/a$ is the Hubble expansion rate and $a(t)$ is the scale factor. The equation of state for each species, with energy densities $\ro_{\rm i}$, and pressures $p_{\rm i}$, take a barotropic form $p_{\rm i}=(\gamma_{\rm i}-1)\ro_{\rm i}$, and the constants $\ga_{\rm i}$ indicate the barotropic index of each component being ${\rm i}=\{x,m,b,r\}$ with $\gamma_{x}=0$, $\ga_{b}=1$ and $\ga_{r}=4/3$, whereas $\ga_{m}=1$. Then $\rho_{x}$ plays the role of a decaying vacuum energy or variable $\Lambda$, $\rho_{m}$ can be associated with dark matter, $\ro_{b}$ represents a pressureless baryonic matter, and $\ro_{r}$ is a radiation fluid. Due to the interaction in the dark sector the conservation equation (\ref{E1b}) cannot be split into two separated components like in equations (\ref{E1c}). In the following it is assumed that there is no interaction between the baryons and the dark sector.

Defining the total dark sector energy density as $\ro_{\dark} \equiv \ro_{m}+\ro_{x}$ together with equation (\ref{E1b}) allows us to express both dark densities as functions of $\ro_{\dark}$ and $\ro'_{\dark}$
\be
\n{04}
\ro_{m}= - \frac{3\ga_{x} \ro_{\dark} +\ro'_{\dark}}{3( \ga_{m}-\ga_{x})}, \qquad \ro_x= \frac{3\ga_{m} \ro_{\dark} +\ro'_{\dark}}{3( \ga_{m}-\ga_{x})},
\ee
where the prime indicates differentiation with respect to the number of e-folds $N=\ln(a/a_0)$ and $'\equiv d/dN = H^{-1}d/dt$; $a_0$ is the present value of the scale factor.

In order to solve for the evolution of the energy density in the dark sector it is necessary to introduce an energy transfer between the two fluids, given by $Q$. It allows us to split the conservation equation (\ref{E1b}) which, following \cite{Luis,Luis7, Sanchez} can be written as
\be
\label{rod3}
{\ro'}_{m}+3 \ga_m\rho_{m}=-3 Q\,,  \qquad {\ro'}_{x}+3 \ga_x\rho_{x}=3 Q\,.
\ee
These equations imply that there is an exchange rate of energy density of $3Q$ in terms of the number of e-folds, which for $Q$ positive goes from the dark matter to the dark energy. The exchange rate of energy density in terms of the cosmological time as considered in \cite{Wands} is $\mathcal{Q}=3HQ$. From Eqs. (\ref{04}) and (\ref{rod3}), the following source equation \cite{Luis} for the energy density $\ro_{\dark}$ of the dark sector is obtained
\be
\n{rod4}
\ro''_{\dark}+3(\ga_{m}+ \ga_{x})\ro'_{\dark} + 9\ga_{m}\ga_{x}\ro_{\dark} =  9Q(\ga_{m}-\ga_{x})\,.
\ee
The strength of the interaction should depend upon the variables of the dark fluid components, and its precise expression will determine the evolution of these components. As shown in \cite{Wands} any dark energy model can be modeled in the background with a suitable choice of the interaction term. This is an example of the dark degeneracy studied in \cite{Kunz:2007rk} and \cite{Aviles:2011ak}.

In the present work we use a generalization from the interaction term in Refs. \cite{Luis7} and \cite{Sanchez}. As we will see, this model tends to a running vacuum $\Lambda(H)$ model with even powers of $H$ that could come from the general covariance of the action of QFT in a curved background \cite{Lima}. The interaction $Q$ between both dark components will be assumed as a non linear interaction given by
\begin{equation}
\label{Q}
Q= \frac{\rho'_{\dark}}{3} \left( \alpha+\beta\rho_{\dark} \right) \,,
\end{equation}
where $\alpha$ and $\beta$ are the coupling constants that measure the strength of the interaction in the dark sector, $\alpha$ being a dimensionless constant while $\beta$ has dimensions of the inverse of the energy density. In another way, the interaction can be written in terms of the physical time as $\mathcal{Q}=3HQ=\dot{\rho}_{\dark} (\alpha + \beta\rho_{\dark})$. Using this expression for the background interaction it's possible to integrate the conservation equation (\ref{rod4}) with the barotropic constants chosen as in the $\Lambda$CDM as
\begin{equation}
	\gamma_m= 1 \,, \qquad \gamma_x =0 \,.
	\label{}
\end{equation}
This choice in the equation of state of the dark energy fluid means that it has an energy momentum tensor proportional to the metric, given by
\begin{equation}
	T_{(x)\mu\nu} = - \rho_x g_{\mu \nu} \,.
	\label{}
\end{equation}
As it is shown in \cite{Wands}, this is the simplest model of dark energy after a cosmological constant, as it doesn't add new degrees of freedom to the dynamics of the model with respect to $\Lambda$CDM. In this type of interaction scenario the dark energy can be viewed as a variable cosmological constant (also known in the specialized literature as running or decaying vacuum models), which depends on time at the background level and on time and space on the next orders in perturbation theory.

Replacing the specific form of $Q$ into the source equation (\ref{rod4}) and the value $\ga_{x}=0$, the  first integral of the second order differential equation for the total energy density $\ro(a)$ of the dark sector is
\begin{equation}
\label{ro1}
3\rho'_{\dark}=\beta\frac{\rho_{\dark}^2}{2}+(\alpha-1)\rho_{\dark}+\mathcal{C},
\end{equation}
where $\mathcal{C}$ is an integration constant. Integrating (\ref{ro1}), under the condition $1 > 2\beta\mathcal{C} /(\alpha-1)^2$, we obtain the energy density of the dark sector as a function of the scale factor $a$:
\be
\label{rod14}
\rho_{\dark} = \frac{(\alpha-1)[1+\mathcal{R} - \mathcal{K}(1-\mathcal{R})a^{-3(\alpha-1)\mathcal{R}}]}{\beta [\mathcal{K}a^{-3(\alpha-1)\mathcal{R}} - 1]},
\ee
where $\mathcal{K}$ is an integration constant and we have defined $\mathcal{R} = \sqrt{1 - 2\beta\mathcal{C} /(\alpha-1)^2}$. We can write the integration constants
$\mathcal{C}$ and $\mathcal{K}$ in terms of the present densities $\rho_{i0}$ as
\begin{eqnarray}
\label{K}
\mathcal{C} &=& \rho_{x0}-\frac{\beta}{2}\rho_{\dark 0}^2-\alpha \rho_{\dark 0}\,, \\
\mathcal{K} &=& \frac{\beta \rho_{\dark 0} + (\alpha-1)(\mathcal{R} + 1)}
{\beta \rho_{\dark 0} - (\alpha-1)(\mathcal{R}-1)}\,.
\end{eqnarray}
The above expressions are valid for $\beta \ne 0$, while for $\beta =0$ the evolution of the density was obtained in \cite{Sanchez} as
\begin{equation}
	\rho_{\dark}= \frac{1}{1-\alpha} ( \rho_{m0}a^{-3(1-\alpha)} + \mathcal{C}),
	\label{rhob0}
\end{equation}
with $\mathcal{C}$ defined as in (\ref{K}). Therefore, the total energy density, or the Friedmann equation (\ref{E1a}), is given by
\be
\ro_{T} = 3{H_0}^2(1-\Omega_{b0}-\Omega_{x0}-\Omega_{m0})a^{-4} + 3{H_0}^2\Omega_{b0}a^{-3}
+ \frac{(\alpha-1)[1+\mathcal{R} - \mathcal{K}(1-\mathcal{R})a^{-3(\alpha-1)\mathcal{R}}]}{\beta [\mathcal{K}a^{-3(\alpha-1)\mathcal{R}} - 1]}. \n{ET}
\ee
Replacing the Eq. (\ref{ro1}) in the second equation of (\ref{04}), we have that the dark energy density as a function of the energy density of the dark sector,
\be
\label{rom}
\ro_{x}=\frac{\beta}{2}\rho_{\dark}^2+\alpha\rho_{\dark}+\mathcal{C},
\ee
and the dark matter density in terms of the total dark sector density is
\begin{equation}
	\rho_m=- \frac{\beta}{2}\rho_{\dark}^2 + (1-\alpha) \rho_{\dark} - \mathcal{C} \,.
	\label{}
\end{equation}
At late times when the dark sector dominates the whole dynamics of the universe the total density can be approximated as $\rho_T\approx\rho$. Under this consideration the Eq. (\ref{rom}) with the proposed interaction model goes to $\Lambda(H)\simeq n_0 + n_1H^2 + n_2H^4$, where $n_0$, $n_1$ and $n_2$ are constants. This particular dependence looks like the one used in decaying vacuum models. It can be shown that under some considerations this limit of the interacting model proposed in this paper boils down into a particular case of a decaying vacuum. To do this we have to consider a decaying vacuum of the form of the limit case of (\ref{rom}) and that only (or mainly) decays into the dominant component, therefore for late times it decays into dark matter \cite{Lima}, \cite{Sola}. By this way we found a link between the interaction models with $\Lambda(t)$ models.

The evolution of the dark fluid density is modified with respect to the dark matter + $\Lambda$ density in the $\Lambda$CDM model in which
\begin{equation}
	\rho_m^{\Lambda CDM} \propto a^{-3} \,,\qquad \rho_x^{\Lambda CDM} = {\rm Const.}
	\label{}
\end{equation}

\begin{figure}[!h]
\centering
\includegraphics[width=0.6\textwidth]{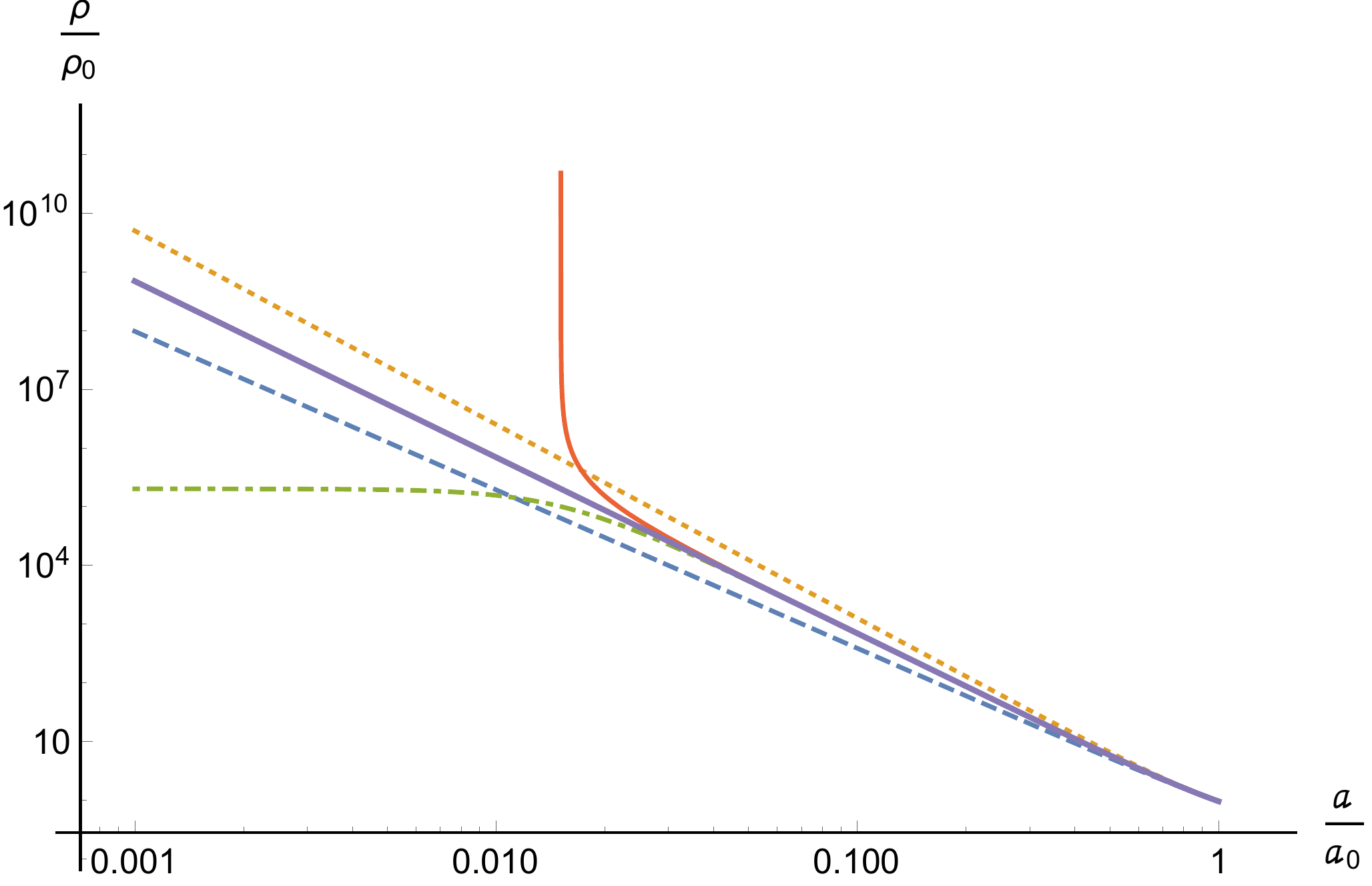}

\caption{Density of the dark fluid as a function of the scale factor. Today's density parameters were fixed to $\Omega_m=0.26$ and $\Omega_x=0.69$. The continuous thick purple line corresponds to the dark energy plus dark matter density in $\Lambda$CDM. The blue dashed and orange doted lines correspond to the model with $\beta=0$ and $\alpha=\pm 0.1$ respectively, we see that the dark fluid density decays slower for $\alpha$ positive and faster for $\alpha$ negative this has an impact on the time of equality between matter and radiation which in turns modifies the position of the matter power spectrum maximum as can be seen in figure \ref{fig:matterpower}. We also generalized this interaction to a perturbation lever. The dot-dashed green line corresponds to $\alpha=0$ and $\beta = (23eV^{-1})^{4}$ while the red continuous line corresponds to $\alpha=0$ and $\beta = - (23eV^{-1})^{4}$. The model with positive $\beta$ tends to a constant value in the past, while the model with negative $\beta$ has a density divergence in the past. This divergence would spoil nucleosynthesis unless it happened much before $a=10^{-10}$ which means that if $\beta$ is negative then $|\beta|\ll 5\times 10^{-20} eV^{-4}$.}
  \label{fig:dark_density}
\end{figure}

This modification is due to the interaction term. In the case in which $\alpha=0$ and $\beta=0$ the $\Lambda$CDM model is recovered. This fact can be seen in Fig. \ref{fig:dark_density}, where we can see that the main effect of $\beta \neq 0$, corresponding to the red continuous line or the green dash doted line, happens for a particular redshift. In fact, it is easy to show that for $\beta<0$, with small absolute values of $\alpha$ and $\rho_{0}\beta$, the density diverges for a redshift of approximately
\begin{equation}
  z_{\rm crit} \approx \left(  \frac{2}{- \beta \rho_{0}} \right) ^{1/3} \,.
  \label{}
\end{equation}
In order to solve this unphysical divergence, we need to make sure that it occurs sufficiently into the past so that it doesn't spoil the big bang nucleosynthesis. In other words $z_{\rm crit} \gg z_{\rm nuc}$ or in terms of $\beta$
\begin{equation}
	|\beta| \ll 5\times 10^{-20} eV^{-4} \,,
  \label{cotab1}
\end{equation}
for $\beta$ negative.
On the other side, for $\beta$ positive the dark density tends to a constant at high redshift (small scale factor). In the limit for $\alpha$ and $\beta\rho_{0}$ small, the density tends to
\begin{equation}
	\rho_{\dark} (a\to 0) \approx \beta^{-1}\,.
  \label{}
\end{equation}
In order to satisfy the constraints on the matter-radiation equality, this density should be bigger than the radiation density at the equality epoch, in other words, for $\beta$ positive it is required that
\begin{equation}
	\beta \ll 0.7 eV^{-4} \,.
  \label{cotab2}
\end{equation}
The preliminary constraint (\ref{cotab2}) will be improved by using CMB + Large Scale Structure (LSS) observations in the section \ref{sec:obs}. The parameter space of $\beta$ negative will be left unexplored due to its small range compared to the $\beta$ positive scale, and the difficulties of treating such a small space using a Markov chain Montecarlo code as we used here.

The evolution of the dark energy component of the dark fluid changes with respect to the constant value in $\Lambda$CDM. As $\rho'$ is negative, we can see from equation (\ref{Q}) that for $\alpha$ and $\beta$ positive, the interaction is negative, corresponding to an energy density exchange from the dark energy to the dark matter. This scenario will alleviate the problem of the smallness of the dark energy density \cite{Copeland:2006wr}. As it can be seen from equation (\ref{rom}), in this scenario the dark energy density can start with a value of the same order of magnitude as the other components of the universe and decrease with time at a lower pace than the other components until it becomes the dominant component as it is today.

\section{Perturbation equations}\label{sec:pert}

The most general scalar perturbed Friedmann-Robertson-Walker metric is given by \cite{Malik:2008im}
\begin{equation}
\label{dsConfEs}
ds^2= (1+2\phi)dt^2 - 2a(t) B_{,i}dx^idt
- a^{2}(t)\left[ (1-2\psi)\delta_{ij} + 2E_{,ij} \right] dx^idx^j \,,
\end{equation}
where $\phi$, $B$, $\psi$ and $E$ are the gauge-dependent scalar perturbation quantities.

The expression (\ref{Q}) for the interaction is valid only at background level. In a more general setting the interaction is a 4-vector which encodes the energy and momentum transfer between two fluids, in this case the dark matter and dark energy. The conservation equation states
\begin{eqnarray}
	\nabla_{\mu} T_{\rm{m}}^{\mu\nu} &=& \mathcal{Q}^\nu \,,\\
	\nabla_{\mu} T_{\rm{x}}^{\mu\nu} &=& - \mathcal{Q}^\nu \,.
	\label{}
\end{eqnarray}
Following ref. \cite{Kodama:1985bj} we can split the interaction in a component along the velocity and another orthogonal to it. In that case
\begin{equation}
	\mathcal{Q}^\nu = \mathcal{Q}_{tot} u^\nu + f^\nu \,.
\end{equation}
The orthogonal component $f$ encodes the (minus) momentum transfer, or force exerted by the dark energy on the dark matter component and can be different from zero only at the perturbation level. The interaction $\mathcal{Q}_{tot}=3HQ + \delta Q$ represents the energy transfer and can be split between a background value given by (\ref{Q}) and a
linear perturbation.

In the linearly perturbed universe, the conservation equations (\ref{rod3}) for the components of dark matter and interacting vacuum energy, reduce to
\begin{eqnarray}
\label{pertrho}
\dot{\delta\rho_{m}}+3H\delta\rho_{m}-3\rho_{m}\dot{\psi}+\rho_{m}\frac{\bigtriangledown^2}{a^2}(\theta_m+a^2\dot{E}-aB)=-\delta Q- \mathcal{Q} \phi, \\
\dot{\delta\rho_x}=\delta Q+ \mathcal{Q} \phi,
\end{eqnarray}
while the momentum conservation becomes
\begin{eqnarray}
\label{pertmom}
\rho_m\dot{\theta}_m+\rho_m\phi=-f=-\mathcal{Q}(\theta_x-\theta_m), \\
-\delta\rho_x= f+ \mathcal{Q} \theta_m \,. \label{pertmom2}
\end{eqnarray}
Combining the equations of (\ref{pertrho}) and the ones of (\ref{pertmom})
we can write
\begin{eqnarray}
\label{pertrho2}
\dot{\delta\rho_{m}}+3H\delta\rho_{m}-3\rho_{m}\dot{\psi}+\rho_{m}\frac{\bigtriangledown^2}{a^2}(\theta_m+a^2\dot{E}-aB)=- \dot{\delta\rho_x}, \\
\rho_m\dot{\theta}_m+\rho_m\phi=\delta\rho_x+ \mathcal{Q}\theta_m.
\end{eqnarray}

We can eliminate two variables choosing the synchronous gauge, $\phi=B=0$. Besides, the metric potentials left can be simplified to $-3\psi+\nabla^2E=h/2$, where $h$ is the trace of the spatial part of the metric and characterizes a scalar mode of spatial metric perturbations. Under these conditions the Eq. (\ref{pertrho2}) takes the form
\begin{equation}
\dot{\delta\rho_{m}}+3H\delta\rho_{m}+\rho_{m}\frac{\dot{h}}{2}
+\rho_m \frac{\nabla^2}{a^2}\theta_m=0.
\end{equation}
In \cite{Wang:2013qy} it is shown that the observations seem to favor a model where $f=0$. Making that choice for our model implies, from equation (\ref{pertmom}) in the synchronous gauge that
\begin{equation}
	\rho_m \dot{\theta}_m=0 \,.
	\label{}
\end{equation}
This equation is equivalent to the corresponding one for the velocity of the dark matter in $\Lambda$CDM. Using an additional gauge freedom for the synchronous gauge, we can choose the initial dark matter velocity to be zero $\theta_m(t_i)=0$ and the above equation will keep the velocity zero for all times as long as the linear approximation remains valid. From eq. (\ref{pertmom2})
it implies that $\delta \rho_x=0$ and the only perturbation from the dark fluid will be that of the matter density which satisfy the equation
\begin{equation}
\label{pertrho4}
\dot{\delta\rho_{m}}+3H\delta\rho_{m}+\rho_{m}\frac{\dot{h}}{2}
=0.
\end{equation}
In order to use the CAMB code \cite{Lewis:1999bs} to compute the evolution of the linear perturbations we make the change of variable $\delta=\delta\rho/\rho$ in Eq. (\ref{pertrho4}), we arrive to
\begin{equation}
\label{pertrho5}
\dot{\delta_{m}}-\mathcal{Q}\frac{\delta_{m}}{\rho_{m}}+\frac{\dot{h}}{2}=0,
\end{equation}
where $\mathcal{Q}=3HQ=\alpha\dot{\rho}+\beta\rho\dot{\rho}$.

\subsection{Growth factor}
The field equation in the synchronous gauge is \cite{Ma}
\begin{equation}
\label{FielSG}
\ddot{h}+2H\dot{h}=-8\pi G(\delta\rho+3\delta P).
\end{equation}
Taking the derivative of Eq. (\ref{pertrho5}) and using Eq. (\ref{FielSG}), we can obtain the second-order differential equations for the dark matter and baryon over-density respectively
\begin{eqnarray}
\label{pertrho6}
\ddot{\delta}_m+\left(2H-\frac{\mathcal{Q}}{\rho_m}\right)\dot{\delta}_m-\left(2H\frac{\mathcal{Q}}{\rho_m}+\frac{\dot{\mathcal{Q}}}{\rho_m}-\frac{\mathcal{Q}\dot{\rho}_m}{\rho_m^2}\right) \delta_m=4\pi G(\rho_m\delta_m+\rho_b\delta_b), \\
\label{pertrho7}
\ddot{\delta}_b+2H\dot{\delta}_b =4\pi G(\rho_m\delta_m+\rho_b\delta_b).
\end{eqnarray}
Replacing the time variable $t$ by $N=\log a/a_0$ and defining the function $g(a)\equiv\delta/a$ we can write the equation system, Eqs. (\ref{pertrho6}) and (\ref{pertrho7}), as

\begin{equation}\label{pertrho8}
g_{m}''+\left[ 3+(\ln \mathcal{H})'-\frac{a\mathcal{Q}}{\mathcal{H}\rho_m}\right] g_m'
+\left[ 2+(\ln \mathcal{H})' - 3\frac{a\mathcal{Q}}{\mathcal{H}\rho_m} - \frac{a\mathcal{Q}'}{\mathcal{H}\rho_m} + \frac{a\mathcal{Q}\rho_m'}{\mathcal{H}\rho_m^2} \right]g_m =  \frac{4\pi Ga^2}{\mathcal{H}^2}(\rho_m g_m + \rho_b g_b),
\end{equation}

\begin{equation}
\label{pertrho9}
g_{b}''+\left[ 3+(\ln \mathcal{H})'\right] g_b'+\left[ 2+(\ln \mathcal{H})' \right]g_b =  \frac{4\pi Ga^2}{\mathcal{H}^2}(\rho_m g_m+\rho_b g_b),
\end{equation}

where $\mathcal{H} = aH$ is the conformal Hubble parameter and the primes now denote the derivations with respect to the variable $N$. We can obtain the evolution of the overall growth rate of matter $f_M$ with redshift $z$ by solving the closed differential system numerically, Eqs. (\ref{pertrho8}) and (\ref{pertrho9}), with the initial conditions in the matter-domination era, $g_i(a_0) = 1$ and $g_i'(a_0) = 0$. We define $f_M$ as
\begin{equation}
f_M\equiv [\ln \delta_M(a)]'=1+g_M'(a),
\end{equation}
where
\begin{equation}
\rho_M g_M(a)=\rho_mg_m+\rho_bg_b,
\end{equation}
and $\rho_M=\rho_m+\rho_b$.

\section{Comparison with observations}\label{sec:obs}

In order to constrain the parameters of the model we use the code COSMOMC \cite{Lewis:2002ah} to sample a parameter space with the 6 usual free parameters from $\Lambda$CDM plus the coupling parameters $\alpha$ and $\beta$ from our model. We use likelihoods from Planck measurements of the CMB \cite{Adam:2015rua}, matter power spectrum measurements from SDSS Luminous Red Galaxies catalog \cite{Eisenstein:2001cq},
and WiggleZ Dark Energy Survey \cite{Blake:2011en}, and BAO measurements from 6dFGRS \cite{Jones:2009yz} and BOSS \cite{Anderson:2013zyy}.

\begin{figure}
\begin{subfigure}{.49\textwidth}
  \centering
  \includegraphics[width=1\textwidth]{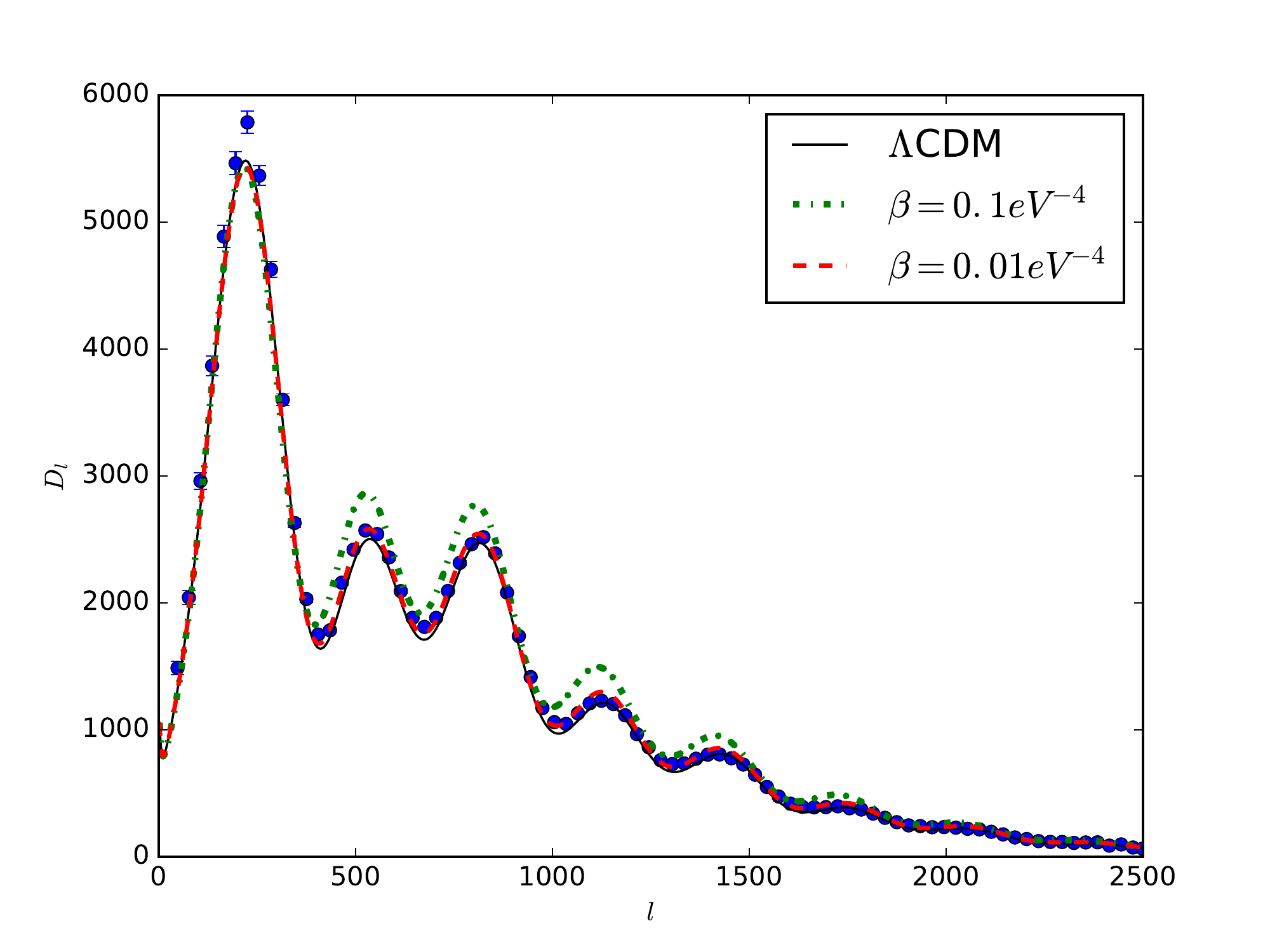}
  \caption{CMB power spectrum for a model with $\alpha=0$ and $\beta$ equal to $0$ ($\Lambda$CDM), $0.01 {\rm eV}^{-4}$ and $0.1 {\rm eV}^{-4}$ respectively.}
  \end{subfigure}\hfill
  \begin{subfigure}{.49\textwidth}
  \centering
  \includegraphics[width=1\textwidth]{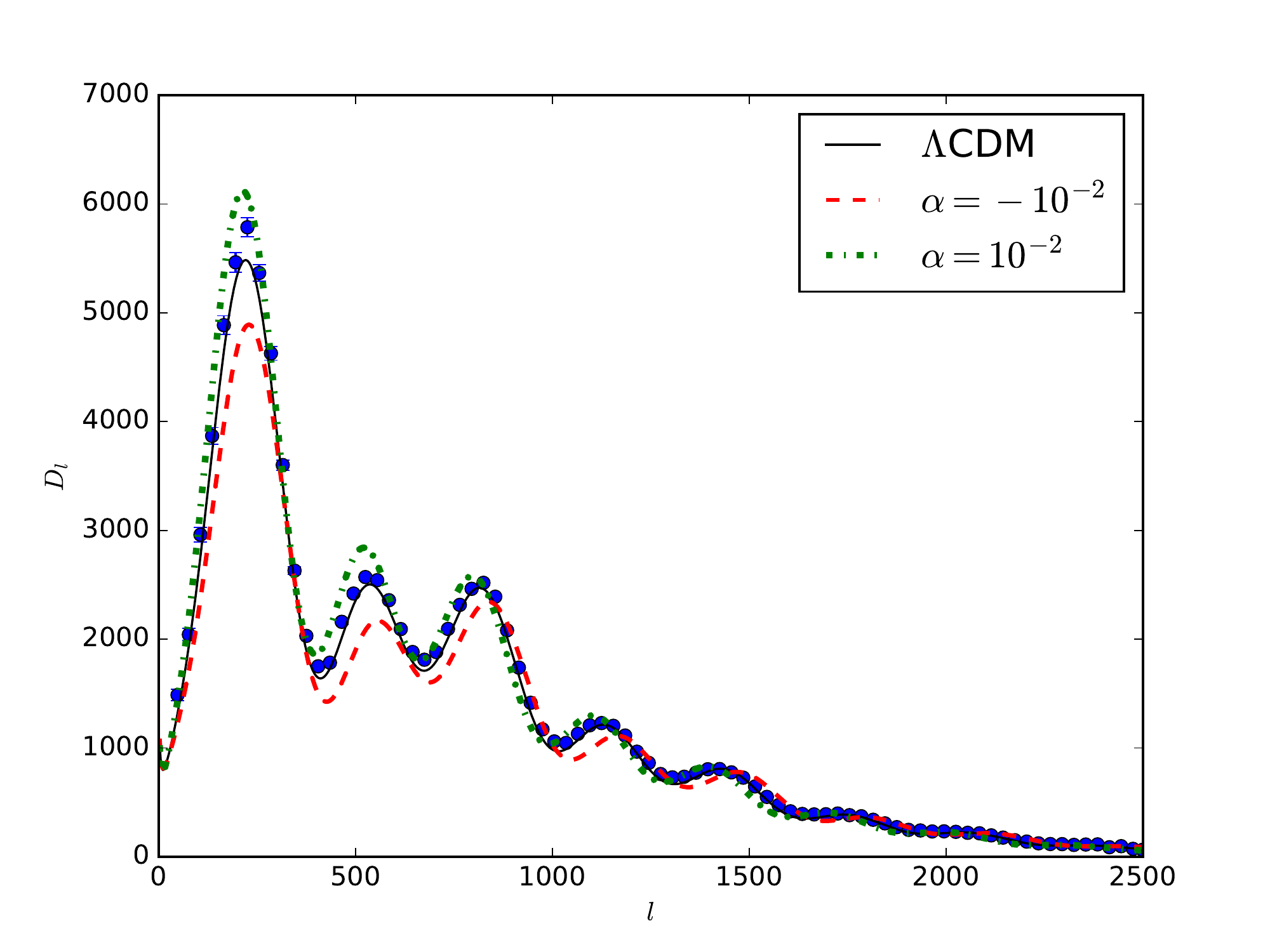}
  \caption{CMB power spectrum for a model with $\beta=0$ and $\alpha$ equal to $-10^{-2}$, $0$ ($\Lambda$CDM) and $10^{-2}$ respectively.}
  \end{subfigure}
  \caption{The CMB temperature power spectrum for different choices of the interaction parameters for comparison we plot also the high $l$ binned multipoles from Planck 2015. The parameters other than the interaction strength are taken from the best fit analysis in Planck's paper \cite{Ade:2015xua}. The parameters $\alpha$ and $\beta$ are taken outside the region of validity obtained in table \ref{tab1} in order to show more clearly the effects of the interaction.} \label{fig:CMB}
\end{figure}

\begin{figure}
\begin{subfigure}{.47\textwidth}
  \centering
  \includegraphics[width=\textwidth]{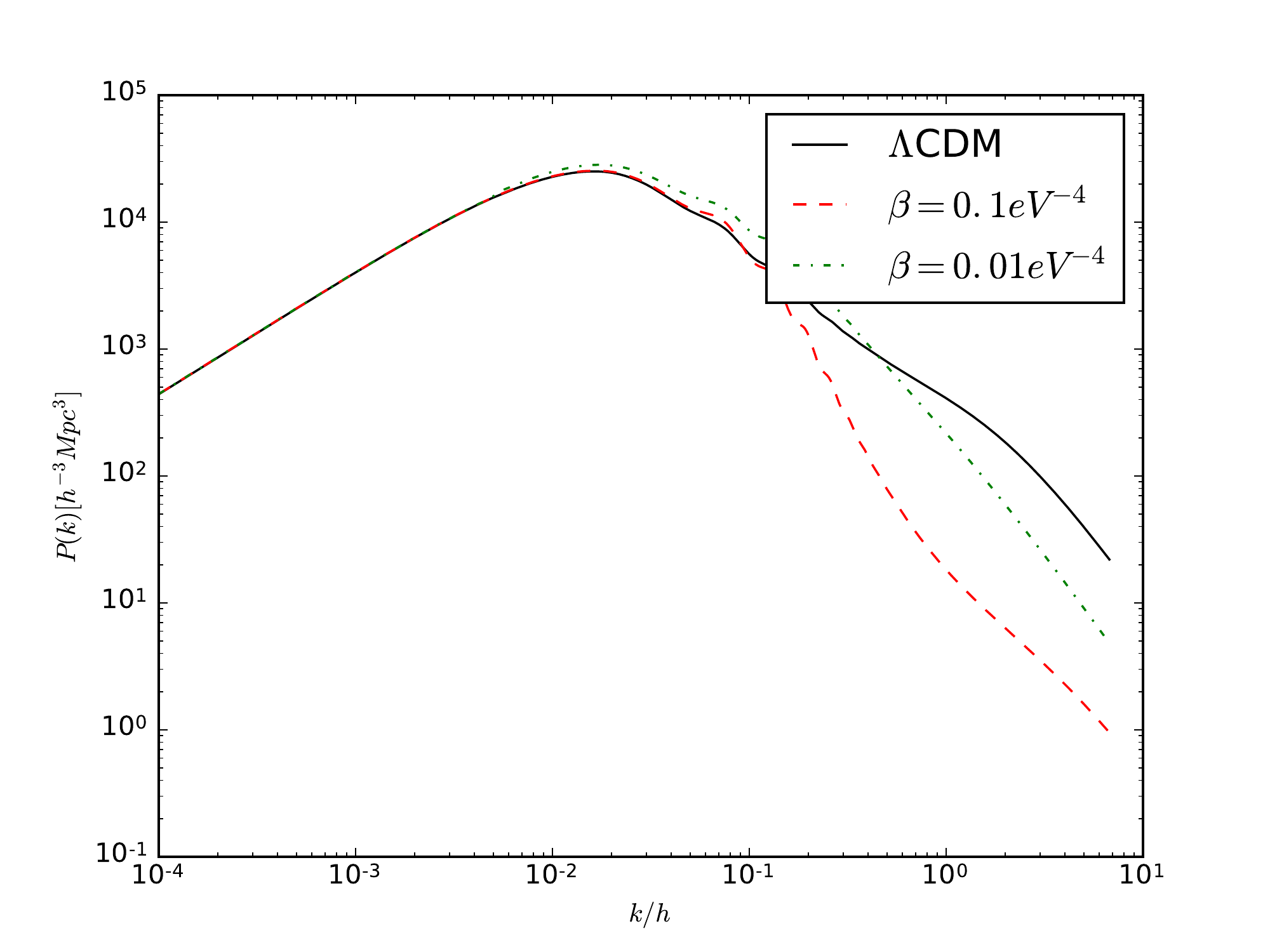}
  \caption{Matter power spectrum for a model with $\alpha=0$ and $\beta$ equal to $0$ ($\Lambda$CDM), $0.01 {\rm eV}^{-4}$ and $0.1 {\rm eV}^{-4}$ respectively.}
  \end{subfigure}\hfill
  \begin{subfigure}{.47\textwidth}
  \centering
  \includegraphics[width=\textwidth]{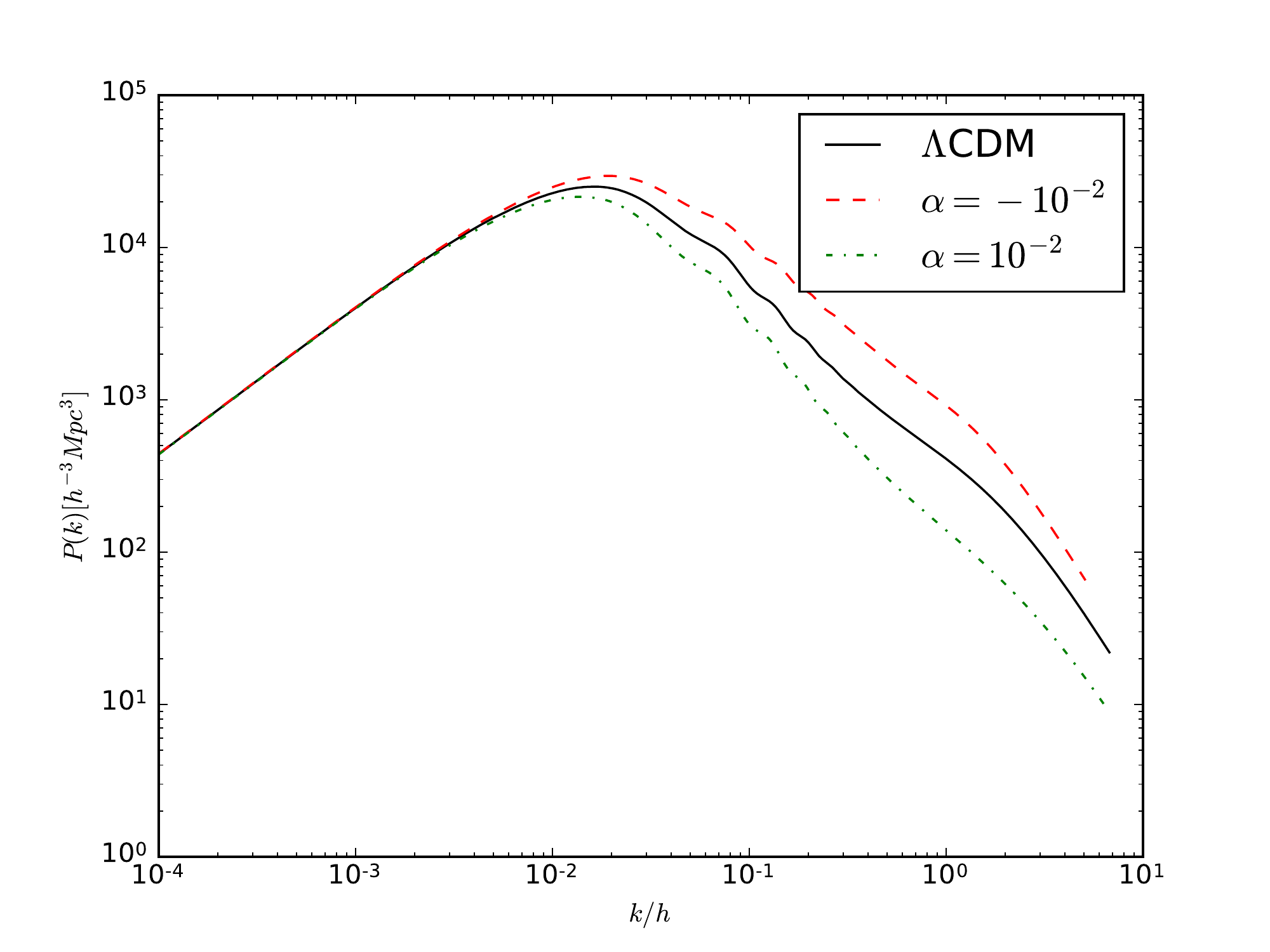}
  \caption{CMB power spectrum for a model with $\beta=0$ and $\alpha$ equal to $-10^{-2}$, $0$ ($\Lambda$CDM) and $10^{-2}$ respectively.}
  \end{subfigure}
  \caption{The matter power spectrum for different choices of the interaction parameters. Power spectrum information were used to compare the model with observations as stated in section \ref{sec:obs}. The parameters $\alpha$ and $\beta$ are taken outside their range of validity in for demonstrative purposes. The comparison with observations leads to the bounds in the parameters written in table \ref{tab1}.}
  \label{fig:matterpower}
\end{figure}

We see in figure \ref{fig:CMB} that the different parameters have different effects in the CMB and matter power spectra. The parameter $\alpha$ rises both the first and second maximums of the spectrum, while $\beta$ acts mainly on the second and third maximums. In figure \ref{fig:matterpower} we can see that the effect of $\beta$ on the matter power spectrum is more significant than on the CMB even for smaller values of the parameter for that reason LSS data will be important to constrain the model.

\begin{figure}[h]
	\begin{subfigure}[t]{.45\textwidth}
  \centering
  \includegraphics[width=0.9\textwidth]{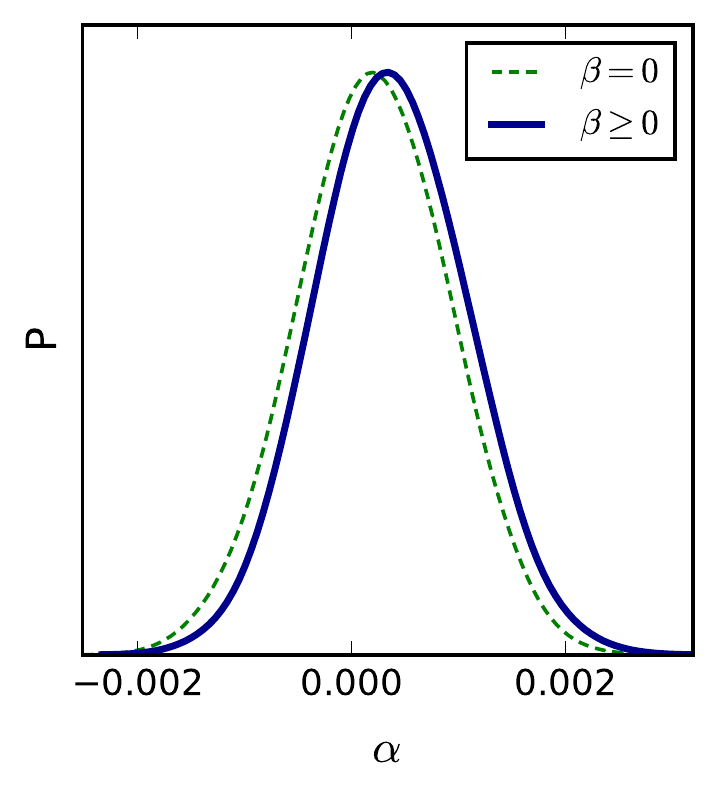}
  \caption{Marginalized posterior distribution for the interaction parameter $\alpha$. The green doted line corresponds to an MCMC run where $\beta=0$ while the blue continuous line corresponds to both, $\alpha$ and $\beta$ allowed to vary. We can see that the presence of $\beta$ has little impact on the posterior distribution of $\alpha$.}
  \end{subfigure}\hfill
  \begin{subfigure}[t]{.45\textwidth}
  \centering
  \includegraphics[width=0.9\textwidth]{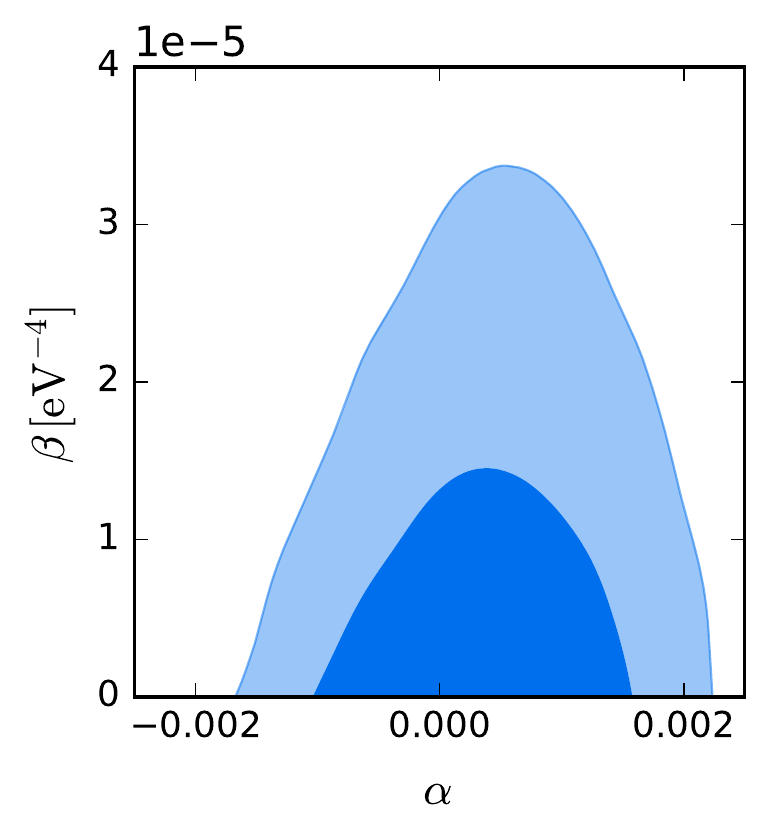}
  \caption{Marginalized posterior distribution for the interaction parameters $\alpha$ and $\beta$ when they are both allowed to vary alongside to the other 6 standard cosmology parameters. The contours represent $68\%$ and $95\%$ confidence.}
  \end{subfigure}
  \caption{Marginalized distributions obtained from two COSMOMC runs considering CMB and mater power spectrum data. In both runs we considered the 6 parameters from standard cosmology plus the interaction parameters. For the first one we allowed $\alpha$ to vary while $\beta$ was set to zero. For the second run we allowed both $\alpha$ and $\beta$ to vary. We see that the correlation between both parameters is small.}
  \label{fig:constrains}
\end{figure}

As a first extension for the model we set $\beta$ to zero as in \cite{Sanchez} and constrained the parameter $\alpha$ together with the usual 6 parameters from $\Lambda$CDM,
then we left both $\alpha$ and $\beta$ free. The results are in table \ref{tab1}. We see that the constrains in $\alpha$ are very similar for both cases. From the fact that the density diverges for negative $\beta$, we choose it to be positive as a prior. Our analysis in section \ref{sec:model} tells us that, in case of $\beta$ to be negative, $\beta$ should be bigger than $-5\times 10^{-20} {\rm eV}^{-4}$ as in equation (\ref{cotab1}). The constrains on the parameters can also be seen in figure \ref{fig:constrains}.

\begin{table}
	\centering
	\begin{tabular}{|c|c|}
		\hline
		$\alpha = (2.0 \pm 7) \times 10^{-4} $ & $\beta= 0$ \\ \hline \hline
		$\alpha = (3.7 \pm 7 )\times 10^{-4} $ &
		$-(1.5\times10^{-5} {\rm eV}^{-1})^{4} \ll \beta < (0.07 {\rm eV}^{-1})^4$ \\  \hline
		
	\end{tabular}
	\caption{Constrains on the interaction parameters from CMB, MPK and BAO measurements. For the first model $\beta$ was set to zero and for the second one $\beta$ was allowed to have only positive values, the stricter bound for negative $\beta$ comes from considerations of nucleosynthesis.}
	\label{tab1}
\end{table}

\section{Discussions}

In this work we have explored a model of the dark sector of the universe in which there is an interaction between the dark matter and the dark energy (\ref{Q}). This interaction contains two terms, the first of them proportional to the rate of change of the total dark fluid and the second, proportional to the product of the time derivative times the density of the dark fluid. As a result, we obtained a modified evolution of the density of the dark fluid given by the explicit expressions (\ref{rod14}) - (\ref{rhob0}) which we plotted in figure \ref{fig:dark_density}. We also studied the linear perturbations of the new model in section \ref{sec:pert} obtaining their evolution equations, which reduced to equation (\ref{pertrho4}) in the synchronous gauge considering zero momentum transfer. We finally compared the results of our model with measurements of the CMB and matter power spectra obtaining the range of validity of the interaction parameters as summarized in table \ref{tab1}.

By introducing the interaction term $Q$ in Eq. (\ref{Q}) we consider the possibility of an energy transfer from the dark matter to the dark energy when it is positive and on the other direction when the parameters are negative. This kind of interacting dark sector models have been of great interest in the last years for its possibility of reducing the tension between the CMB and RSD measurements as in Ref. \cite{Salvatelli:2014zta}, also, they can explain the smallness of the dark energy density by considering that some of its initial energy was transferred to the dark matter in the past. Finally there is no observational reason to consider the dark matter and dark energy as two separate components as we know very little about their nature. In the literature there are a series of studies which try to unify both phenomena as the result of a single model \cite{Copeland:2006wr}, \cite{Lazkoz:2015mwa}. Indeed, we see through our work, that is much better to consider the dark sector as a single fluid for our analysis in the background, but keeping the freedom of the interaction term to produce non adiabatic perturbations which result in a zero sound speed for the dark matter component which seems to be better suited to reproduce the observations as obtained in \cite{Wang:2013qy}.

In particular the interaction terms used in our work have been studied in \cite{Sanchez} and \cite{Luis7}, with the advantage that in our case we obtain an explicit expression for the full evolution of the dark fluid density as a function of the scale factor and that we also developed an analysis of the perturbations of the model, which allow us to compare the model with a range of different observations of the CMB and matter power spectrum from Planck data to SDSS. In particular, we obtain that the interaction term is consistent with zero, corresponding to the $\Lambda$CDM model, but the mean value corresponds to both $\alpha$ and $\beta$ positive, as summarized in table \ref{tab1} corresponding to an energy transfer from the dark energy to the dark matter.

The question of whether the dark energy is a cosmological constant or some other dynamical component is expected to be settled to a high degree with the new generation of observations in the incoming decade like Euclid \cite{Amendola:2012ys} and DESI \cite{Levi:2013gra}. In this sense, the study of possible modifications to this behavior becomes important today. The building of this theoretic constructs will pay of in the next years when we will be able to test the nature of dark energy. The interactive models seems promising as they are a simple generalization from de $\Lambda$CDM model that would be tested in the near future.

\section*{Acknowledgments}
We gratefully acknowledge ICTP for the organization and support of the First ICTP Advanced School on Cosmology
where discussions for this work started.
JDS acknowledge posdoctoral grant from DGAPA-UNAM at ICF and SNI-CONACYT for support. ISG acknowledge posdoctoral from CONICET at IFIBA and DF FCEN-UBA for support. DT acknowledge posdoctoral grant from CONACYT at CINVESTAV.
\vskip 1cm



\begin{thebibliography}{99}

\bibitem{Riess1}
A. G. Riess, \emph{et al.}, Astrophys. J. \textbf{116} (1998) 1009.

\bibitem{Perlm}
S. Perlmutter, \emph{et al.}, Astrophys. J. \textbf{517} (1999) 565.

\bibitem{Adam:2015rua} 
R. Adam, \emph{et al.}, Astron. Astrophys. \textbf{594} (2016) A1.

\bibitem{BAO}
K. S. Dawson, \emph{et al.}, Astronomical Journal \textbf{145} (2013) 10 

\bibitem{Adelman}
J. K. Adelman-McCarthy, \emph{et al.}, Astrophys. J. Suppl. \textbf{175} (2008) 297.

\bibitem{Drees}
M. Drees and G. Gerbier, arXiv:1204.2373 (2012).

\bibitem{Garrett}
K. Garrett and G. Duda, Adv. Astron. \textbf{2011} (2011) 968283.

\bibitem{WMAP}
G. Hinshaw, G. \textit{et al.}, Astrophys. J. Suppl. \textbf{208} (2013) 19.

\bibitem{Planck1}
P. A. R. Ade \emph{et al.}, Astron. Astrophys. \textbf{571} (2014) A15.

\bibitem{Planck2}
P. A. R. Ade \emph{et al.}, Astron. Astrophys. \textbf{571} (2014) A16.

\bibitem{Weinberg1989}
S. Weinberg. Rev. Mod. Phys., \textbf{61} (1989) 1.

\bibitem{Grande}
J. Grande \emph{et al.}, JCAP \textbf{1108} (2011) 007.

\bibitem{Alcaniz}
J. S. Alcaniz \emph{et al.}, Phys. Lett. B \textbf{716} (2012) 165.

\bibitem{Sola}
J. Sol{\`a}, J. Phys. Conf. Ser. \textbf{453} (2013) 012015.

\bibitem{Perico} 
  E. L. D. Perico, J. A. S. Lima, S. Basilakos and J. Sol{\`a}, Phys. Rev. D {\bf 88} (2013) 063531.
  
\bibitem{Gomez-Valent}
J. Sol{\`a}, A. G{\'o}mez-Valent, J. Cruz P{\'e}rez, Astrophys. J. Lett. \textbf{811} (2015) L14, (2015).

\bibitem{Tamayo}
D. A. Tamayo, J. A. S. Lima, M. E. S. Alves, and J. C. N. de Araujo, Astropart. Phys. \textbf{87} (2017) 18.

\bibitem{Perico:2016kbu}
E. L. D. Perico and D. A. Tamayo, JCAP \textbf{1708} (2017) 026.

\bibitem{Ferreira} 
P.G. Ferreira and M. Joyce, Phys. Rev. D \textbf{58} (1998) 023503.
 
\bibitem{Stein}
P. J. Steinhardt, L. M. Wang and I. Zlatev, Phys. Rev. D \textbf{59} (1999) 123504. 

\bibitem{Sahni}
V. Sahni and L.M. Wang, Phys. Rev. D \textbf{62} (2000) 103517.

\bibitem{Luis}
L. P. Chimento, Phys. Rev. D \textbf{81} (2010) 043525.

\bibitem{Amendola1}
L. Amendola, Phys. Rev. Lett. \textbf{86} (2001) 196.

\bibitem{Amendola2}
L. Amendola, Phys. Rev. D \textbf{69} (2004) 103524.

\bibitem{Zimdahl}
W. Zimdahl, D. Pav\'{o}n, L. P. Chimento,  Phys. Lett. B \textbf{521} (2001) 133.

\bibitem{Koivisto}
T. Koivisto, Phys. Rev. D \textbf{72} (2005) 043516.

\bibitem{Ziaeepour}
H. Ziaeepour, Phys. Rev. D \textbf{86} (2012) 043503.

\bibitem{Luis7}
L. P. Chimento, M. G. Richarte and I. E. S\'{a}nchez G., Phys. Rev. D \textbf{88} (2013) 087301.

\bibitem{Sanchez}
I. E. S\'{a}nchez G., Gen. Rel. and Grav. \textbf{46} (2014) 1769.

\bibitem{Marachlian}
E. Marachlian, I. E. S\'{a}nchez G., O. Santill\'{a}n, arXiv:1508.05083 (2015).

\bibitem{Pavon}
S. del Campo, R. Herrera and D. Pav\'{o}n,  Phys. Rev. D \textbf{91} (2015) 123539.

\bibitem{Vali}
J. V{\"a}liviita and E. Palmgren, JCAP \textbf{1507} (2015) 015.

\bibitem{Riess3}
A. G. Riess \textit{et al.}, Astrophys. J. \textbf{730} (2011) 119.

\bibitem{Planck3}
P. A. R. Ade \emph{et al.}, Astron. Astrophys. \textbf{571} (2014) A20.

\bibitem{Wang}
Y. Wang, D. Wands, G. Zhao and L. Xu, Phys. Rev. D. \textbf{90} (2014) 023502.

\bibitem{Murgia}
R. Murgia, S. Gariazzo and N. Fornengo, JCAP \textbf{1604} (2016) 014.

\bibitem{Lewis:2002ah} 
A. Lewis and S. Bridle, Phys. Rev. D {\bf 66} (2002) 103511.

\bibitem{Wands}
D. Wands, J. De-Santiago, Y. Wang, Class. Quantum Grav. \textbf{29} (2012) 145017.

\bibitem{Kunz:2007rk}
M. Kunz, Phys. Rev. D {\bf 80} (2009) 123001.

\bibitem{Aviles:2011ak}
A. Aviles and J. L. Cervantes-Cota, Phys. Rev. D {\bf 84} (2011) 083515. [Erratum: Phys. Rev. D {\bf 84} (2011) 089905].

\bibitem{Lima}
J. A. S. Lima, S. Basilakos and J. Sol{\`a}, Mon. Not. Roy. Astron. Soc. \textbf{431} (2013) 923.

\bibitem{Copeland:2006wr}
E. J. Copeland, M. Sami and S. Tsujikawa, Int. J. Mod. Phys. D {\bf 15} (2006) 1753.

\bibitem{Malik:2008im}
K. A. Malik and D. Wands, Phys. Rept. {\bf 475} (2009) 1.

\bibitem{Kodama:1985bj} 
H. Kodama and M. Sasaki, Prog. Theor. Phys. Suppl. {\bf 78} (1984) 1.

\bibitem{Wang:2013qy} 
Y. Wang, D. Wands, L. Xu, J. De-Santiago and A. Hojjati, Phys. Rev. D {\bf 87} (2013) 083503.

\bibitem{Lewis:1999bs} 
A. Lewis, A. Challinor and A. Lasenby, Astrophys. J. {\bf 538} (2000) 473.

\bibitem{Ma}
C. P. Ma and E. Bertschinger, Astrophys. J. \textbf{455} (1995) 7.



\bibitem{Eisenstein:2001cq} 
D. J. Eisenstein \emph{et al.}, Astron. J. {\bf 122} (2001) 2267.

\bibitem{Blake:2011en} 
C. Blake \emph{et al.}, Mon. Not. Roy. Astron. Soc. {\bf 418} (2011) 1707.

\bibitem{Jones:2009yz} 
D. H. Jones \emph{et al.}, Mon. Not. Roy. Astron. Soc. {\bf 399} (2009) 683.

\bibitem{Anderson:2013zyy} 
L. Anderson \emph{et al.}, Mon. Not. Roy. Astron. Soc. {\bf 441} (2014) 24.

\bibitem{Ade:2015xua} 
P. A. R. Ade \emph{et al.}, Astron. Astrophys. {\bf 594} (2016) A13.

\bibitem{Salvatelli:2014zta}
V. Salvatelli, N. Said, M. Bruni, A. Melchiorri and D. Wands, Phys. Rev. Lett. {\bf 113} (2014) 181301.

\bibitem{Lazkoz:2015mwa}
R. Lazkoz, I. Leanizbarrutia and V. Salzano, J. Phys. Conf.\ Ser. {\bf 600} (2015) 012028.

\bibitem{Amendola:2012ys}
L. Amendola \emph{et al.}, Living Rev. Rel. {\bf 16} (2013) 6.

\bibitem{Levi:2013gra}
M. Levi \emph{et al.}, arXiv:1308.0847 (2013).
 
\end{thebibliography}
\end{document}